%% file: main.tex
\documentclass[12pt]{article}

\input{preamble.tex}

\input{commands.tex}

\begin{document}
\thispagestyle{empty}
\begin{center}
{\titlesize\bf Experience-driven formation of parts-based representations in a model of layered visual memory}
\end{center}

\vspace{1mm}

\begin{center}
{\bf Jenia Jitsev$^{1,2, *}$ and Christoph von der Malsburg$^{1}$ } 
\end{center}
\vspace{-3,5mm}
{$^1$Frankfurt Institute of Advanced Studies, Frankfurt Am Main, Germany \\
 $^2$Johann Wolfgang Goethe University, Frankfurt Am Main, Germany \\}

\vspace{5mm}
	
{\bf \hspace{-7,5mm} Running Title: \\}
Formation of layered visual memory. \\

\vspace{5mm}
	
{\bf \hspace{-7,5mm} Correspondence: \\}
Jenia Jitsev \\
Frankfurt Institute for Advanced Studies (FIAS), \\
Ruth-Moufang-Str.1, 60438 Frankfurt am Main, Germany \\
jitsev@fias.uni-frankfurt.de






\newpage
\thispagestyle{empty}
\include{abstract}

\clearpage
\pagenumbering{arabic} 

\include{introduction}

\include{materials_methods}

\include{results}

\include{discussion}

\include{acknowledge}






\bibliography{main_bib}
\bibliographystyle{elsarticle-harv}


\end{document}

%% file: preamble.tex
\usepackage[latin1]{inputenc}
\usepackage[T1]{fontenc}
\usepackage{times}
\usepackage[pdftex]{graphicx}
\usepackage{amsmath}
\usepackage{amssymb}
\usepackage{natbib}

\usepackage{titlesec} 
\usepackage{subfigure}



%% file: commands.tex
\newcommand{\Norm}[2]{\ensuremath{\lVert #1 \rVert_{\!_{\hspace{1pt} #2}}}}
\newcommand{\NormTwo}[1]{\Norm{#1}{2}}

\newcommand{\VectorP}{\ensuremath{\vec{\mathbf{p}}}}

\DeclareMathOperator*{\maxOp}{max}

\titleformat{\section}{\bfseries}{\thesection. \quad}{1em}{}
\titleformat{\subsection}{\bfseries}{\thesubsection.}{1em}{}

\newcommand{\titlesize}{\fontsize{16pt}{20pt}\selectfont} 


%% file: abstract.tex
\section*{Abstract}
\label{sec:abstract}
Growing neuropsychological and neurophysiological evidence suggests that the visual cortex uses parts-based representations to encode, store and retrieve relevant objects. In such a scheme, objects are represented as a set of spatially distributed local features, or parts, arranged in stereotypical fashion. To encode the local appearance and to represent the relations between the constituent parts, there has to be an appropriate memory structure formed by previous experience with visual objects. Here, we propose a model how a hierarchical memory structure supporting efficient storage and rapid recall of parts-based representations can be established by an experience-driven process of self-organization. The process is based on the collaboration of slow bidirectional synaptic plasticity and homeostatic unit activity regulation, both running at the top of fast activity dynamics with winner-take-all character modulated by an oscillatory rhythm. These neural mechanisms lay down the basis for cooperation and competition between the distributed units and their synaptic connections. Choosing human face recognition as a test task, we show that, under the condition of open-ended, unsupervised incremental learning, the system is able to form memory traces for individual faces in a parts-based fashion. On a lower memory layer the synaptic structure is developed to represent local facial features and their interrelations, while the identities of different persons are captured explicitly on a higher layer. An additional property of the resulting representations is the sparseness of both the activity during the recall and the synaptic patterns comprising the memory traces.

\section*{Keywords}
\label{sec:keywords}
Visual memory, self-organization, unsupervised learning, competitive learning, bidirectional plasticity, activity homeostasis, parts-based representation, cortical column

%% file: introduction.tex
\section{Introduction}
\label{sec:introduction}
A working hypothesis of cognitive neuroscience states that the higher functions of the brain require coordinated interplay of multiple cortical areas distributed over the brain-wide network. For instance, the mechanisms of memory are thought to be subserved by various cortical and subcortical regions, including the medial temporal lobe (MTL), inferior temporal (IT) and prefrontal (PFC) cortex areas \citep{Fuster1997, Miyashita2004} to name only few of them prominent in the function of the visual memory. Studies of information processing going on in the course of encoding, consolidation and retrieval of visual representations reveal a hierarchical organization, sparse distributed activity and massive recurrent communication within the memory structure \citep{Tsao2006, Konen2008, Osada2008}. Here we focus our attention on developmental issues and discuss the process of self-organization that may lead to the formation of the core structure responsible for flexible, rapid and efficient memory function, with organizational properties as inferred from the experimental works. 

It is widely held that processes responsible for memory formation rely on activity-dependent modification of the synaptic transmission and on regulation of the intrinsic properties of single neurons \citep{Miyashita1988, Bear1996, Zhang2003}. However, it is far from clear how these local processes could be orchestrated for memorizing complex visual objects composed of many spatially distributed subparts arranged in stereotypic relations. In mature cortex, there is strong evidence for a basic vocabulary of shape primitives and elementary object parts in the TEO and TE areas of posterior and anterior IT \citep{Fujita1992, Tanaka2003} as well as for identity and category specific neurons in anterior IT, PFC and hippocampus \citep{Freedman2003, Quiroga2005}. Further findings indicate that the encoding of visual objects involves the formation of sparse clusters of distributed activity across the processing hierarchy within inferior temporal cortex \citep{Tsunoda2001,Reddy2006a}. This seems to be a neuronal basis for the parts-based representation that the visual system employs to construct objects from their constituent part elements \citep{Ullman2002, Hayworth2006}.

In the light of these findings, we may ask ourselves whether the observed memory organization happens to be the outcome of a self-organization process that would have to find solution to a number of developmental tasks. To provide a neural substrate for the parts-based representation, memory traces have to be formed and maintained in an unsupervised fashion to span the basic vocabulary for the visual elements and to define associative links between them. Subsets of associatively linked complex features can then be interpreted as coherent objects composed of the respective parts. As there is a virtually unlimited number of visual objects in the environment, the limited resources spent on formation of these memory traces have to be carefully allocated to avoid unfavorable interference effects and information loss caused by potential memory content overlap. Thus, the system is permanently confronted with the problem of selecting the right small population out of the totally available, potentially conflicting synaptic facilities which has to be modified for acquisition and consolidation of a novel stimulus. Moreover, if objects stored in memory are supposed to share common parts, a regulation mechanism would be required to balance the usage load of part-specific units and minimize the interference, reassuring their optimal participation in memory content formation and encoding. Another issue is the timing of the modifications, which have to be coordinated properly if the correct relational structure of distributed parts constituting the object's identity is to be stored in the memory.

The same selection problem arises on the fast time scale, during memory recall or for encoding of a novel object. Currently, there is a broad agreement on the sparseness of the activity patterns evoked by the presentation of a complex visual object, where only a small fraction of the available neurons in the higher visual cortex participate in the stimulus-related response \citep{Rolls1995, Olshausen2004, Quiroga2008}. In the context of the parts-based representation scheme, one possible interpretation of sparse activation would be the selection of few parts from a large overcomplete vocabulary for the composition of the global visual object. Considering the speed of object recognition measured in psychophysical experiments on humans and primates \citep{Thorpe2001a}, there have to be neural mechanisms allowing this selection procedure to happen within the very short time of a few hundred milliseconds. Moreover, if relations are to be represented by dynamic assemblies of co-activated part-specific neurons, such a combinatorial selection would require clear unambiguous temporal correlations between the constituent neurons to identify them and only them as being part of the same assembly encoding the object \citep{CvdM99, Singer1999a}.

Hypothesizing that the process of neural resource selection and its coordination across distributed units is a crucial ingredient for successful structure formation and learning, we address in this study the neural mechanisms behind the selection process by incorporating them in a model of a layered visual memory. Here we take the competition and cooperation between the neuronal units as the functional basis for the structure formation \citep{Malsburg1988, Edelman1993} and provide modification mechanisms based on activity-dependent bidirectional plasticity \citep{Bienenstock1982, Artola1993} and homeostatic activity regulation \citep{Desai1999b}. We confront the system with a task of unsupervised learning and human face recognition using a database of natural face images. Our aim is then to demonstrate the formation of synaptic memory structure comprising bottom-up, lateral and top-down connectivity. 

Starting from an initial undifferentiated connectivity state, the system is able to form a representational basis for the storage of individual faces in a parts-based fashion by developing memory traces for each individual person over repetitive presentations of the face images. The memory traces are residing in the scaffold of lateral and top-down connectivity making up the content of the associative memory that holds the associatively linked local features on the lower and the configurational global identity on the higher memory layer. The recognition of face identity can then be explicitly signaled by the units on the higher memory layer (Fig. \ref{fig:VMM}). By performing this self-organization, the system solves a highly non-trivial and important problem of capturing simultaneously local and global signal structure in an unsupervised, open-ended fashion, learning not only the appearance of local parts, but also memorizing their combinations to represent the global stimulus identity explicitly in lateral and top-down connectivity. None of the previous works on unsupervised learning of natural object representation were able to solve this problem in this explicit form \citep{Waydo2008, Wallis2008}.

As a consequence of this explicit representation, the local facial features are interpreted in the global context of the identity of a person, making use of the structure formed in the course of previous experience. This contextual structure can also be utilized in generative fashion to replay the memory content in absence of external stimuli, also supporting the mechanism of selective object-based attention. The binding of the local features and their identity label into a coherent assembly is done in the course of a decision cycle spanned by a common oscillatory rhythm. The rhythm  modulates the competition strength and builds up a frame for repetitive local winner-take-all computation. As the agreement between incoming bottom-up, lateral and top-down signals gets continuously improved during the competitive learning, the bound assemblies tend to reflect more and more consistently the face identities stored in the memory, so that the recognition error progressively decreases. Moreover, the employment of the contextual connectivity speeds up the learning progress and leads to a greater capability to generalize over novel data not shown before. The advanced view on the structure formation as an optimization process driven by evolutionary mechanisms of selection and amplification may also serve as a conceptual basis for studying self-organization of generic subsystem coordination, independent of the nature of the cognitive task.

%% file: materials_methods.tex
\section{Materials and Methods}
\label{sec:methods}

\subsection{Visual memory network organization}
\label{subsec:network}
\begin{figure}[!tbhp]
\includegraphics[scale=1.0]{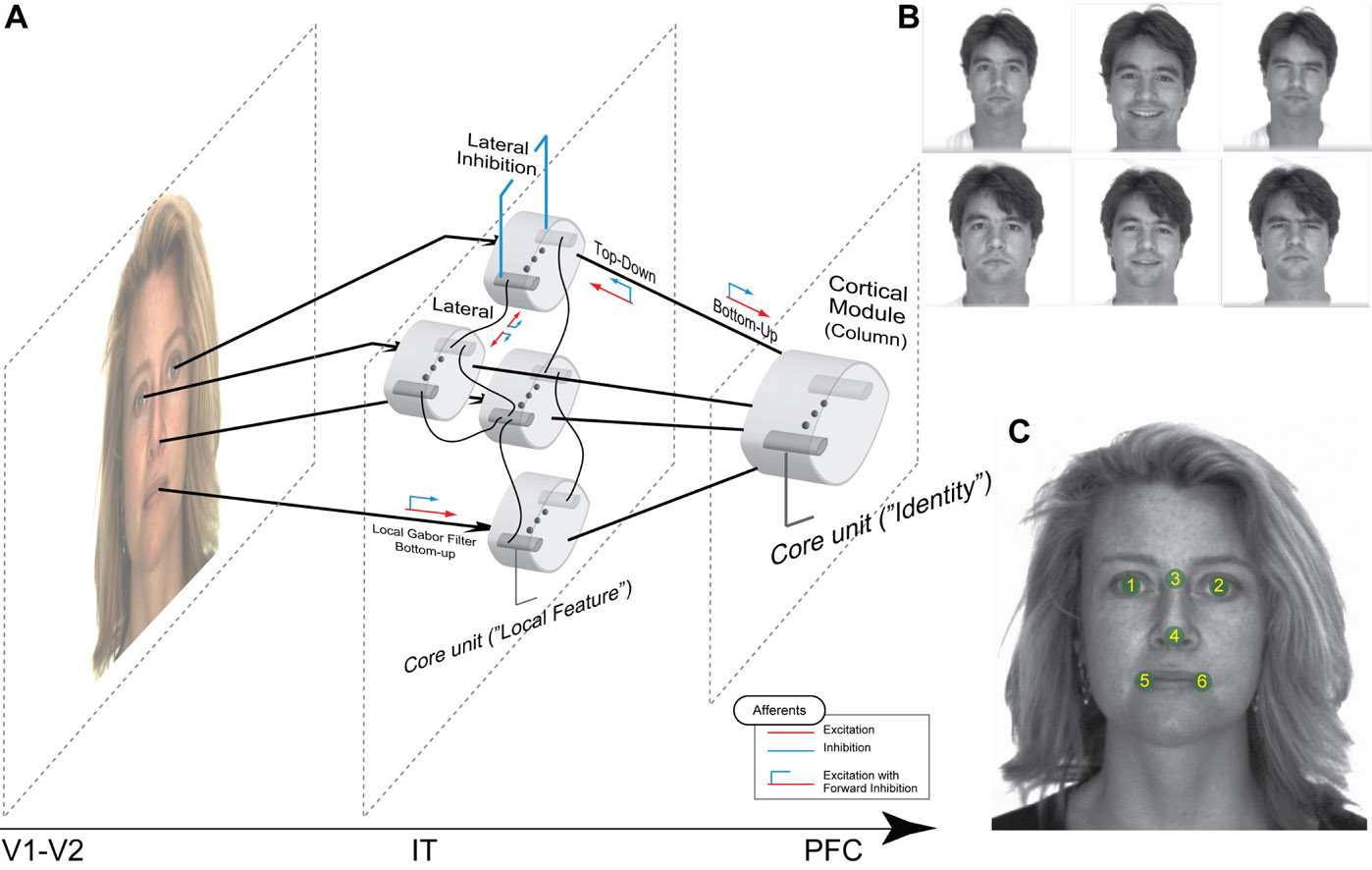}
\caption[Layered Visual Memory]{Layered visual memory model. {\bf (A)} Two consecutive interconnected layers for hierarchical processing. On the lower bunch layer (IT, each column contains $n=20$ units), a storehouse of local parts linked associatively via lateral connections is formed by unsupervised learning. On the higher identity layer (PFC, column contains $m=40$ units), symbols for person identities emerge, being semantically rooted in parts-based representations of the lower layer. The identity units provide further contextual support for the lower layer by establishing top-down projections to the corresponding part-specific units. {\bf(B)} Different face views used as input to the memory (one person out of total $40$ used for learning shown). Top left is the original view with neutral expression used for learning. Other views were used for testing the generalization performance (bottom row shows the duplicate views taken two weeks after the original series.). {\bf(C)} Facial landmarks used for the sensory input to the memory, provided by Gabor filter banks extracted at each landmark point}
\label{fig:VMM}
\end{figure}

Our model is based on two consecutive interconnected layers (Fig. \ref{fig:VMM}), which we tend to identify with the hierarchically organized regions of IT and PFC, containing a number of segregated cortical modules that will be termed {\it columns} \citep{Fujita1992, Mountcastle1997a, Tanaka2003}. The columns situated on the lower layer will be termed here {\it bunch} columns, as each of them are supposed to hold a set of local facial features acquired in the course of learning. The column on the higher memory layer will be called {\it identity} column as its task will be to learn the global face identity for each individual person composed out of distributed local features on the lower memory layer. Being a local processing module, each column contains further a number of subunits we call {\it core units} (or simply {\it units}), which receive common excitatory afferents and are bound by common lateral inhibition. Acting as elementary processing units of the network, the core units represent an analogy to a tightly coupled population of excitatory pyramidal neurons (``pyramidal core'') as documented in cortical layers II/III and V \citep{PetersCifuentesSethares1997, Rockland2004, Yoshimura2005a}. These populations are thought to be capable of sustaining their own activity even if afferent drive is removed.

On the lower level of processing, each bunch column is attached to a dedicated landmark on the face to process the sensory signal represented by a Gabor filter bank extracted locally from the image \citep{Daugman1985, CvdM97}. The connections bunch units receive from the image constitute their bottom-up receptive fields (here, referring to a receptive field we always mean the pattern of synaptic connections converging on a unit). Furthermore, there are excitatory lateral connections between the bunch columns on the lower layer binding the core units across the modules. The bunch units also send bottom-up efferents to and get top-down afferent projections from the identity units situated on the higher level of processing. All the types of intercolumnar synapses are excitatory and plastic, the connectivity structure being all-to-all homogeneous in the initial state.

\subsection{Dynamics of a core unit}
\label{subsec:core}
A cortical column module containing a set of $n$ core units is modeled by a set of $n$ differential equations each describing the dynamic behavior of the unit's activity variable $p$. The basic form of the equation, ignoring the afferent inputs for the time being, is motivated by a previous computational study on a cortical column \citep{Lucke2005}:
\begin{equation}
\label{eq:basicGL}
\tau \dfrac{dp}{dt} =  \alpha p^2 (1-p) - \beta p^3 - \lambda \nu (\maxOp(\VectorP_{t}) - p) p ,
\end{equation}
where $\tau$ is the time constant, $\alpha$ the strength of the self-excitability, $\beta$ the strength of self-inhibitory effects, $\lambda$ the strength of the lateral inhibition between the units, $\nu$ the inhibitory oscillation signal and $\maxOp(\VectorP_{t})$ the activity of the strongest unit in the column module. In this study we set for all units $\tau = 0.02\,ms$, $\alpha = \beta = 1$, $\lambda = 2$. As $p$ reflects the activity of a whole neuronal population receiving common afferents, we may assume a small time constant value, referring to an almost instantaneous response behavior of a sufficiently large ($n=100$ or more) population of neurons \citep{Gerstner2000}.

\begin{figure}[!htbp]
\includegraphics[scale=1.0]{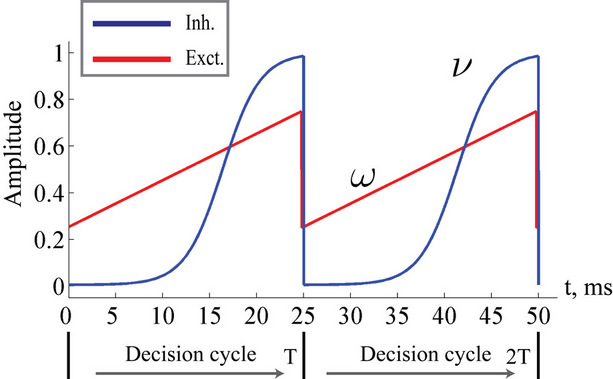}
\caption[Oscillations]{Excitatory ($\omega$) and inhibitory ($\nu$) oscillation rhythms defining a decision cycle in the gamma range.}
\label{fig:oscillation}
\end{figure}

A crucial property of the column dynamics is the ability to change the structure of the stable activity states by variation of the parameter $\nu$. We take the oscillatory inhibition activity $\nu$ (Fig. \ref{fig:oscillation}) to be of a form
\begin{equation}
\label{eq:inhibOscl}
\nu(t) =  \nu_{min} + \frac{1}{k \cdot e^{-g\,(mod(t,T) \,-\, 0.5\,(T + T_{init}))} \,+\, (\nu_{max} - \nu_{min})^{-1}}
\end{equation}
with its period $T = 25\, ms$ being in the gamma range. $\nu_{min}$ and $\nu_{max}$ are the lower and upper bounds for oscillation amplitude, $T_{init}$, $k$, $g$ parameterize the form of the sigmoid activity curve. Here the values are set to $\nu_{min} = 0.005$, $\nu_{max} = 1.0$, $T_{init} = 5ms$, $g = 0.5$, $k = 2$. With the rising strength of the oscillatory inhibition, the parameter $\nu$ crosses a critical bifurcation point of structural instability which is given by:
\begin{equation}
\label{eq:nuCritic}
\nu_c = \frac{\alpha}{\lambda}
\end{equation}
so that by inserting the given values of $\alpha$ and $\lambda$ we obtain $\nu_c = 0.5$. For the range $\nu < \nu_c$ any units subset can remain active (with the stationary activity level $p = \tfrac{\alpha}{\alpha + \beta}$), as these states are stable given the low strength of lateral inhibition. After crossing the critical value $\nu_c$, all those states having more than one unit active loose the stability, so that only a single winner unit can remain active on the level $\tfrac{\alpha}{\alpha + \beta}$. The bifurcation property realizes winner-take-all behavior of the column acting as a competitive decision unit \citep{Lucke2005} to select the best response alternative on the basis of the incoming input.

The qualitative dynamical behavior stays the same in the extended formulation of the activity equation, which is:
\begin{equation}
\begin{aligned}\label{eq:fullGL}
\tau \dfrac{dp}{dt} = \quad &  \alpha \omega (1 + \kappa^{LAT} I^{LAT} + \kappa^{TD} I^{TD}) p^2 (1-p) - \beta p^3 \\ 
		      &  - \lambda \omega \nu (\maxOp(\VectorP_{t}) - p) p + \kappa^{BU} I^{BU} p^2 + \theta p \\
		      &  + \omega \epsilon + \sigma \eta_{t} p ,
\end{aligned}
\end{equation}
where $I^{BU}$, $I^{LAT}$, $I^{TD}$ are the afferent inputs of respective bottom-up, lateral and top-down origin, $\kappa^{BU} = \kappa^{LAT}$ = $\kappa^{TD} = 1$ are their coupling coefficients, $\omega$ is an excitatory oscillatory signal, $\theta$ an excitability threshold of the unit, $\sigma = 0.001$ is parameterizing the multiplicative gaussian white noise $\eta_{t}$ and $\epsilon$ is an unspecific excitatory drive. $\theta$ is a dynamic threshold variable used for homeostatic activity regulation of the unit, it will be described later in detail; $\epsilon$ depends on the total number of core units $n$, $\epsilon = \tfrac{1}{5n}$.

An important modeling assumption is the separation of the synapses of different origin as implemented in Eq.\ref{eq:fullGL}. This separation causes different synaptic inputs to have different impact on the activity of the unit. The functional difference can be made explicitly evident by taking a glance at the stable state of the winner unit (assuming for clarity $\sigma = \epsilon = \theta = 0$), which takes the value
\begin{equation}
\label{eq:pStable}
p_{stable} = \dfrac{\alpha \omega (1 + \kappa^{LAT} I^{LAT} + \kappa^{TD} I^{TD}) + \kappa^{BU} I^{BU}}{\alpha \omega (1 + \kappa^{LAT} I^{LAT} + \kappa^{TD} I^{TD}) + \beta} ,
\end{equation}
where bottom-up input $I^{BU}$ contributes to the activity level in a linear fashion, while the contribution of lateral and top-down inputs $I^{LAT}$ and $I^{TD}$ is non-linear, resembling the pure driving and hybrid driving-modulating roles of afferents from different origin commonly assumed for cortical processing \citep{Sherman1998, Friston2005}. The course of the activity is also influenced by the excitatory oscillatory activity $\omega$ (Fig. \ref{fig:oscillation}), which is given by:
\begin{equation}
\label{eq:excOscl}
\omega(t) = \omega_{min} + \frac{mod(t,T)}{T}\, (\omega_{max} - \omega_{min}) ,
\end{equation}
where $\omega_{min} = 0.25$ and $\omega_{max} = 0.75$ are the lower and upper bounds for oscillation amplitude. The excitatory oscillation doesn't make any impact on the critical bifurcation point $\nu_c$, as it modulates the self-excitability strength $\alpha$ and the lateral inhibition strength $\lambda$ to the same extent (Eq. \ref{eq:fullGL}). Instead, it elevates the activity level of the units as long as they manage to resist the rising inhibition and remain in the active state. In the state where lateral inhibition gets strong enough to shut down all but the strongest core unit, only this winner unit is affected by the elevating impact of the excitatory oscillation, being able to further amplify its activity at the cost of suppressing the others. Both inhibitory and excitatory oscillations may have presumably different sources, the former being generated by the interneuron network of fast-spiking (FS) inhibitory cells \citep{Whittington1995} and the latter having its origin in activities of fast rhythmic bursting (FRB), or chattering, excitatory neurons \citep{Gray1996}.

In addition to the local competitive mechanism supported by the lateral inhibition within a column, we use a simple form of forward inhibition \citep{Douglas1991} acting on the incoming afferents. To model this, the incoming presynaptic activities are transformed as following before they make up the afferent input via the respective receptive field of a unit:
\begin{equation}
\label{eq:forwardInh}
\begin{aligned}
\hat{p}^{pre}_{i} = p^{pre}_{i} - {\displaystyle \dfrac{1}{K} \sum_{j}^{K} p^{pre}_{j}}, & \qquad pre \in \lbrace BU, LAT, TD \rbrace \\
I^{Source} = {\displaystyle \sum_{i}^{K} w^{Source}_{i} \hat{p}^{pre}_{i}}, & \qquad Source \in \lbrace BU, LAT, TD \rbrace ,
\end{aligned}
\end{equation}
where $p^{pre}$ stands for raw presynaptic activity, $\hat{p}^{pre}$ is the presynaptic activity transformed by forward inhibition, $K$ is the total number of incoming synapses of a certain origin, the weights $w^{Source}_{i}$ constitute the receptive field and $I^{Source}$ designates the final computed value of the afferent input from the respective origin. Although all plastic synaptic connections in the network are taken to be of excitatory nature, the forward inhibition allows units to exert inhibitory action across the columns. An important effect of this processing is the selection and amplification of strong incoming activities at the cost of weaker ones, which can be interpreted as presynaptic competition among the afferent signals \citep{Douglas1991, Swadlow2003}.

An additional property of the dynamics is the natural restriction of the population activity values $p$ to the interval between $0$ and $1$ (Eq. \ref{eq:pStable}), given that the afferent input also stays in the same range. This allows both interpretations of the variable as either the population rate or the probability of an arbitrary neuron from the population to generate a spike.

\subsection{Homeostatic activity regulation}
\label{subsec:homeostasis}
The activity dynamics equation (Eq. \ref{eq:fullGL}) contains the variable threshold $\theta$, which regulates the excitability of the unit. Here, higher values of $\theta$ stand for higher unit excitability, implying a greater potential to become active given a certain amount of input. The threshold is updated according to the following rule:
\begin{equation}
\label{eq:homeostatic}
\dfrac{d\theta}{dt} =  \tau_{\theta} (p_{aim} - <p>) ,
\end{equation}
where $<p> = \tfrac{1}{T}{\displaystyle \int_{t}^{t+T} p(t) dt}$ is the average activity of the unit measured over the period $T$ of a decision cycle, $p_{aim}$ specifies the target activity level and $\tau$ is the inverse time constant ($\tau_{\theta} = 10^{-4} ms^{-1}$). The target activity level $p_{aim}$ depends on the number of units $n$ in a column, $p_{aim} = \tfrac{1}{n}$. The initial value of the excitability threshold is zero, $\theta(0) = 0$.

The motivation behind this homeostatic regulation of unit's activity \citep{Desai1999b, Zhang2003} is to encourage a uniform usage load across units in the network, so that their participation on the formation of the memory traces is balanced. Bearing in mind the strongly competitive character of the columnar dynamics, the regulation of the excitability threshold changes the a-priori probability of a unit to be winner of a decision cycle. So, if a certain unit happens to take part too frequently in encoding of the memory content, violating the requirement of the uniform win probability across the units, its excitability will be downregulated so that the core unit becomes more difficult to activate, giving an opportunity for other units to participate in the representation. Reversely, a unit being silent for too long is upregulated, so it can get excited more easily and contribute to memory formation.

\subsection{Activity-dependent bidirectional plasticity}
\label{subsec:plasticity}
We choose a bidirectional modification rule to specify how a synapse connecting one core unit to another may undergo a change in its strength $w$:
\begin{equation}
\label{eq:plasticity}
\dfrac{dw}{dt} =   \varepsilon p^{pre} p^{post} \mathcal{H}(\chi -A(t)) \mathcal{H}(p^{post} - \theta_{0}^{-}) \mathcal{H}_{-}^{+}(p^{post} - \theta_{-}^{+})
\end{equation}
with the sign switch functions $\mathcal{H}(x)$ and $\mathcal{H}_{-}^{+}(x)$ given as following
\begin{equation}\label{eq:heavyside}
\begin{aligned}
& \mathcal{H}(x) =
\begin{cases}
1, \quad x \geq 0 \\
0, \quad x < 0
\end{cases}, 
& \qquad
\mathcal{H}_{-}^{+}(x) =
\begin{cases}
1, \quad x \geq 0 \\
-1, \quad x < 0
\end{cases}
\end{aligned}
\end{equation}
providing the bidirectional form of the synaptic modification. The amplitude of the change is determined by the correlation between the presynaptic activity $p^{pre}$ and the postsynaptic activity $p^{post}$, both variables being non-negative due to the properties of the unit activity dynamics. The learning rate $\varepsilon = 5 \cdot 10^{-4} \,ms^{-1}$ specifies the speed of modification being the inverse time constant. Other variables determine the sign of the modification. The threshold $\theta_{-}^{+} =  \maxOp(\VectorP_{t}^{post})$ is used to compare the postsynaptic activity against current maximum activity in the column. $A(t)$ is the the total activity level in the postsynaptic column at time point $t$, $A(t) = \sum_{i = 1}^{n} p_{i}(t)$, where $n$ is the number of units in the column and $p_{i}(t)$ their activities at time point $t$. $A(t)$ is compared to a variable gating threshold $\chi$, which pursues the average total activity level $<A(t)>$ computed over the period $T$ of a decision cycle:
\begin{equation}\label{eq:chi}
\begin{aligned}
\dfrac{d\chi}{dt} = \tau_{\chi} (<A(t)> - \chi), & \quad <A(t)> = \frac{1}{T}{\displaystyle \int_{t}^{t+T} A(t) dt}
\end{aligned}
\end{equation}
with $\tau_{\chi}=10^{-3}\,ms^{-1}$ as inverse time constant, the threshold initial value set to $\chi(0) = 0.5$. Furthermore, the postsynaptic activity $p^{post}$ is compared to the sliding threshold $\theta_{0}^{-}$ that follows the average postsynaptic activity $<p^{post}(t)>$ computed over the period $T$ of a decision cycle:
\begin{equation}\label{eq:thetaLearning}
\begin{aligned}
\frac{d\theta_{0}^{-}}{dt} = \tau_{\theta_{0}^{-}} (<p^{post}(t)> - \theta_{0}^{-}), & \quad <p^{post}(t)> = \frac{1}{T}{\displaystyle \int_{t}^{t+T} p^{post}(t) dt}
\end{aligned}
\end{equation}
with the inverse time constant $\tau_{\theta_{0}^{-}}=2 \cdot 10^{-3}\,ms^{-1}$, the initial value of the threshold  $\theta_{0}^{-}(0) = p_{aim}$ being equal to the target postsynaptic activity level (see Eq. \ref{eq:homeostatic}).

\begin{figure}[!htbp]
\includegraphics[scale=1.0]{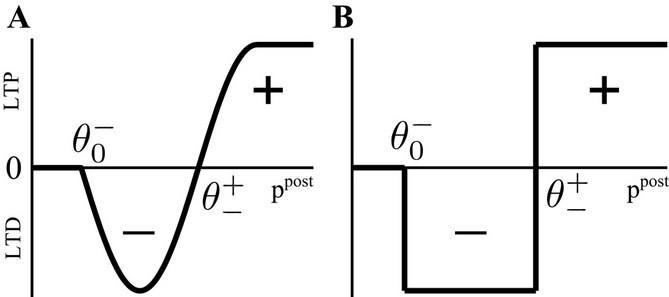}
\caption[Plasticity rule]{Bidirectional plasticity. {\bf (A)} Experimentally grounded modification rule (ABS,  \citealp{Artola1993}) {\bf(B)} A simplified sign switch rule used in the model.}
\label{fig:plasticity}
\end{figure}

The rule employed here is a simplified version of a bidirectional modification assuming the existence of two sliding thresholds $\theta_{0}^{-}$ and $\theta_{-}^{+}$ (Fig. \ref{fig:plasticity}), which subdivide the range of postsynaptic activity into zones where no modification, depression or potentiation may occur, resembling BCM and ABS learning rules rooted in neurophysiological findings \citep{Bienenstock1982,Artola1993,Bear1996, Cho2001}. If the postsynaptic activity level is too low ($p^{post} < \theta_{0}^{-}$), no modification can be triggered. A mediocre level of activation ($\theta_{0}^{-} < p^{post} < \theta_{-}^{+}$) promotes long-term depression (LTD, negative sign), and a high level of activity ($p^{post} > \theta_{-}^{+}$) makes long-term potentiation (LTP, positive sign) possible. Combined with the winner-take-all-like behavior of the core units, the intended effect of the rule is to introduce the competition in synaptic formation across the receptive fields of the units, enabling them to separate patterns even if they are highly similar and overlap strongly. If multiple core units are frequently co-activated by a stimulus, the winner unit gets an advantage in potentiating its stimulated synapses, while the stimulated synapses of the units with lower activity either do not change or are affected by the depression. If this situation occurs over and over, the receptive fields of previously co-activated units are supposed to drift apart preferring the structure where strong synapses are not in conflict with each other anymore, emphasizing the discriminative features of the patterns preferred by the units.

In addition, we here use multiplicative synaptic scaling applied to synapses grouped according to their origin (bottom-up, lateral and top-down). We model this simply by $L^2$-normalization of the receptive field vector, $\tilde{w}^{Source}_{i} =  {w}^{Source}_{i} \big/ \NormTwo{\mathbf{w}^{Source}}$, with ${w}^{Source}_{i}$ as a weight of the receptive field comprising the synapses of the respective origin $Source \in \lbrace BU, LAT, TD \rbrace$, and $\tilde{w}^{Source}_{i}$ its normalized version. The normalizing procedure can be applied after a number of decision cycles, here we choose this number to be $10$ cycles. The scaling mechanism promotes competition between synapses within the receptive field, as the growth of one synapse happens at the cost of the weakening the others \citep{Miller1994a}.

\subsection{Open-ended unsupervised learning and performance evaluation}
\label{subsec:learning}

{\bf Data format.}\hspace{2mm} To provide the system with natural image input, we choose the AR database containing grayscale human face photographs of $126$ persons in total \citep{Martinez1998}. For each person, there is a number of views taken under different conditions (Fig. \ref{fig:VMM} B). The original view with neutral facial expression is accompanied by a duplicate view depicting the same person at a later time point (two weeks after the original shot). Furthermore, there are variations in emotional expression such as smiling or sad for both original and duplicate views. The images were automatically prelabeled with a graph structure put upon the face, positioning nodes on consistent landmarks across different individuals with a software (EAGLE) based on the algorithm described in \citep{CvdM97}. A subset of $L = 6$ facial landmarks was selected around the eyes, nose and mouth regions (Fig. \ref{fig:VMM} C), each landmark being subserved by a single bunch column. Being attached to a dedicated facial landmark, each bunch column is provided with a sensory image signal represented by a Gabor filter bank extracted locally. The Gabor wavelet family used for the filter operation is parameterized by the frequency $k$ and orientation $\varphi$ of the sinusoidal wave and the width of the gaussian envelope $\sigma$  \citep{Daugman1985}. We use $s=5$ different frequencies and $r=8$ different orientations sampled uniformly to construct the full filter bank (for more details refer to \citep{CvdM97}). The local filtering of the image produces a complex vector of responses, containing both amplitude and phase information. We use only the amplitude part consisting of $s \cdot r = 40$ real coefficients to model the responses of complex cells. This amplitude vector is further normalized by $L^2$-Norm to serve as bottom-up input for the respective landmark bunch column of the lower memory layer.

{\bf Network configurations.}\hspace{2mm} Selecting randomly $P = 40$ persons from a database, we allocate $n=20$ core units for each bunch column to ensure that multiple persons have to share some common parts. The identity column then contains $m=40$ units corresponding to the number of persons we want be able to recall explicitly. Two different configurations of the memory system are employed to test our hypothesis about the functional advantage of a fully recurrent structure over the purely feed-forward one. Each configuration is supposed to form the memory structure in the course of the learning phase. While the fully recurrent configuration learns bottom-up, lateral and top-down connectivity, the purely feed-forward configuration is a stripped-off version using only the bottom-up pathways. Observing these different configurations during the learning phase and testing them on novel face views subsequently, we are able to compare both in terms of learning progress and performance on the recognition task to find out potential functional differences between them.

{\bf Simulation.} In order to run the memory network, the solutions for the differential equations governing the behavior of dynamical variables have to be computed numerically in an iterative fashion. We use a simple Euler method with a fixed time step $\Delta t = 0.02 ms$ to do this. To save computational time, slow threshold variables are updated once in a decision cycle, correcting the time steps accordingly.

{\bf Open-ended unsupervised learning.}\hspace{2mm} The system starts with homogeneously initialized structure parameters, all threshold values and all synaptic weights being undifferentiated, so that intercolumnar all-to-all connectivity is the initial structure of the memory network. During the iterative learning procedure, for each decision cycle a face image is selected from a database randomly and presented to the system, evoking a pattern of activity on both memory layers and triggering synaptic and threshold modification mechanisms. The learning procedure is {\it open-ended} as there is neither a stop condition nor an explicitly defined time-dependent learning rate variables which would decrease with time progress and freeze modifications at some point. The learning progress can be assessed directly by evaluating the recognition error on the basis of the previous network responses. Further, the inspection of the structure of the receptive fields delivers hints about their maturation progress. Investigating the rate of ongoing modifications of the synaptic weights and dynamic thresholds could give a hint on whether the changes in the network structure are still taking place in significant proportion, providing a basis for a stop condition if necessary. In the later learning phase the general stability of the established structure can be also verified by simple visual inspection.

{\bf Performance evaluation.}\hspace{2mm} To assess the recognition performance of the system, we make a distinction between the learning and generalization error. The learning error is defined as a rate of wrong responses to person identity from the training data set containing the original face views with neutral expression. The statistics of response behavior to each particular person is gathered for each identity core unit over the history of the network stimulation. The learning error rate can then be computed for each small interval during the learning phase by using the preferences the identity units have developed for the individual persons during the preceding stimulation. Opposed to this, the generalization error is computed on the set of novel views not presented before. During the test for generalization error, all the synaptic weights are frozen, which is done to exclude the possibility that recognition rate improves during the testing phase due to potential benefit of synaptic modifications. The generalization error is assessed for each view type separately to see potential performance differences between different views (the duplicate view and the views with two different emotional expressions, smiling and sad). The history of network behavior during the learning phase is used again in the same way for the computation of the error rate, as done for the learning error evaluation.

\subsection{Assessing network's organization}
\label{subsec:assess}
To analyze the progress of structure formation, we use measures describing different properties of the receptive fields. The distance measure calculates the distance between two synaptic weight vectors $\mathbf{w}_{i}$ and $\mathbf{w}_{j}$:
\begin{equation}\label{eq:distRF}
\begin{aligned}
d(\mathbf{w}_{i}, \mathbf{w}_{j}) & = \frac{1}{4}\, \Bigl(\frac{\mathbf{w}_{i}}{\NormTwo{\mathbf{w}_{i}}} - \frac{\mathbf{w}_{j}}{\NormTwo{\mathbf{w}_{j}}}\Bigr)^{2} \\
& = \frac{1}{2} (1 - \cos\phi) ,
\end{aligned}
\end{equation}
where $\phi$ denotes the angle between the two synaptic weight vectors each comprising a receptive field. The value lies in the interval between zero and one. If the weight vectors are the same, the distance value is zero, if their dissimilarity is maximal ($\alpha = \pi$), the value is one. Utilizing this basic distance measure, we further construct a differentiation measure, which is supposed to reflect the grade of differentiation between the receptive fields of the same type across the whole network. The differentiation grade ${D}^{Source}_{k}$ is computed for each column for the receptive fields of a given type $Source \in \lbrace BU, LAT, TD \rbrace$ and then an average differentiation value ${D}^{Source}$ is built from the values of all $K$ columns:
\begin{equation}\label{eq:diff}
\begin{aligned}
\mathcal{D}^{Source}_{k} = & \frac{1}{n(n-1)}\displaystyle{\sum_{i}^{n} \sum_{j\neq i}^{n} d(\mathbf{w}^{Source}_{i}, \mathbf{w}^{Source}_{j})} \\
\mathcal{D}^{Source} = & \frac{1}{K}\displaystyle{\sum_{k}^{K} {D}^{Source}_{k}} ,
\end{aligned}
\end{equation}
where $n$ is the number of units in the column. The differentiation grade measure is evaluated separately for bunch columns on the lower memory layer and for the identity column on the higher memory layer.

Further we employ a measure reflecting the property of the inner structure of a receptive field to be sparse, that is, possessing few strong synapses and many weak synapses comprising the receptive field. If the inner receptive field structure is poorly differentiated the sparseness value will be low; if differentiation within the receptive field is strong, then the value will be high. To assess the same property not only within, but also across receptive fields, the overlap measure is defined. If the receptive fields of the same type have many strong overlapping synapses in common the value will be high, if there are only few such overlapping synapses the value will be low. The overlap measure is thus closely related to the differentiation grade between the receptive fields as assessed using the distance measure. Both sparseness denoted as $\zeta$ and overlap denoted as $\xi$ have the same scheme behind their computation, with the only difference that the former is computed within while the latter across the receptive field vectors using a common selectivity measure $\mathcal{A}^{Source}(s)$ as defined in \citep{Rolls1995}. Again, the computation is done for each column on receptive fields of the same type $Source \in \lbrace BU, LAT, TD \rbrace$, building then type-specific average values $\mathcal{C}^{Source}$ and $\mathcal{E}^{Source}$ over all $K$ columns:
\begin{equation}\label{eq:sparse_overlap}
\begin{aligned}
& \mathcal{A}^{Source}(s) = \Bigl(\tfrac{1}{s}\sum_{i}^{s} w^{Source}_i\Bigr)^2 \bigg/ \Bigl(\tfrac{1}{s}\sum_{i}^{s} (w^{Source}_{i})^2 \Bigr) \\
& \zeta^{Source}_{k} = \frac{1}{n}\displaystyle{\sum_{i}^{n} \mathcal{A}^{Source}_{i}(r)}, \qquad \xi^{Source}_{k} = \frac{1}{r}\displaystyle{\sum_{i}^{r} (1 - \mathcal{A}^{Source}_{i}(n))} \\
& \mathcal{C}^{Source} = \frac{1}{K}\displaystyle{\sum_{k}^{K} {\zeta}^{Source}_{k}}, \qquad \mathcal{E}^{Source} = \frac{1}{K}\displaystyle{\sum_{k}^{K} {\xi}^{Source}_{k}} ,
\end{aligned}
\end{equation}
where $r$ is the number of synapses comprising a receptive field of type $Source \in \lbrace BU, LAT, TD \rbrace$, $n$ is the number of units in a column, and $K$ is the total number of assessed columns. The evaluation is done separately for the bunch columns and the identity column.

%% file: results.tex
\section{Results}
\label{sec:results}

\subsection{Structure formation}
\label{subsec:formation}
Facing a task of unsupervised learning, the system develops a structural basis for storing the faces of individual persons shown during the learning phase. The vocabularies for the distributed local features are created on the lower memory layer to represent facial parts. These vocabularies are formed by the bottom-up synaptic connections of the bunch columns attached to their facial landmarks. Each core unit of the bunch columns becomes thus sensitive to a particular local facial appearance due to the established structure of its bottom-up receptive field. At the same time, the lateral connectivity between the bunch columns gets shaped capturing the associative relations between the distributed features. These relations are represented by associative links between those core units that are regularly used in the composition of a particular individual face. The same configurational information enters into the structure of bottom-up connectivity converging on the identity column units, being also represented in the top-down connections projecting from the identity column back on the lower layer.

\begin{figure}[!tbhp]
\includegraphics[scale=1.0]{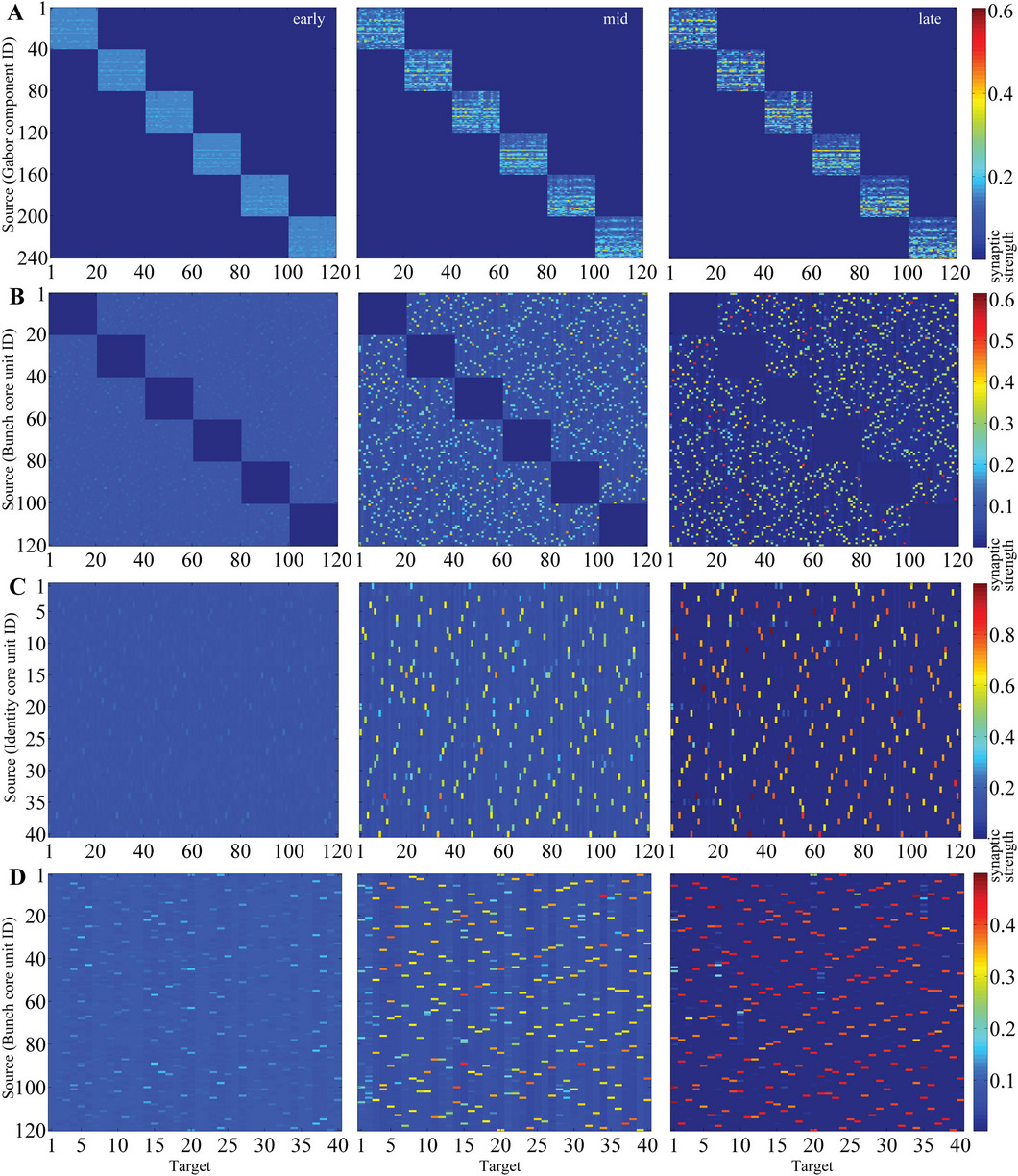}
\caption[Receptive Field Formation]{Time snapshots of structure formation. From left to right, snapshots from early, middle and late formation phase of {\bf (A)} lower layer bottom-up connectivity containing local facial parts, {\bf(B)} lower layer associative lateral connectivity, {\bf (C)} top-down compositional connectivity projecting from the higher back on the lower layer, which is roughly the transposed version of the higher layer bottom-up connectivity visualized in {\bf(D)}, holding global identities.}
\label{fig:RFFormation}
\end{figure}

\begin{figure}[!tbhp]
\includegraphics[scale=1.0]{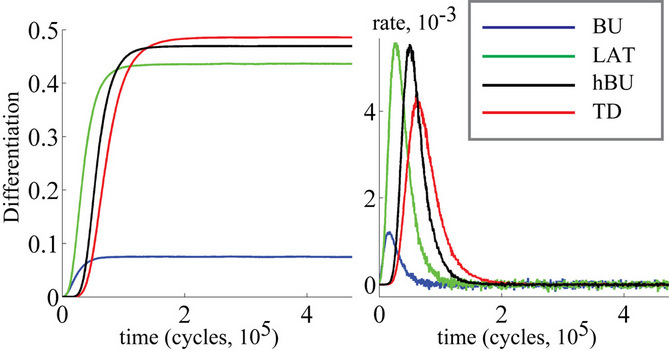}
\caption[Differentiation course]{Differentiation time course over $5 \cdot 10^5$ decision cycles for different connectivity types; on the left the grade of differentiation, on the right its rate. Clear is the general tendency to greater connectivity differentiation with the learning progress as well as the temporal sequence of connectivity maturation (see the text). BU, LAT, hBU, TD denote respectively lower layer bottom-up, lateral, higher layer bottom-up and top-down connectivity types.}
\label{fig:RFDiff}
\end{figure}

\begin{figure}[!tbhp]
\includegraphics[scale=1.0]{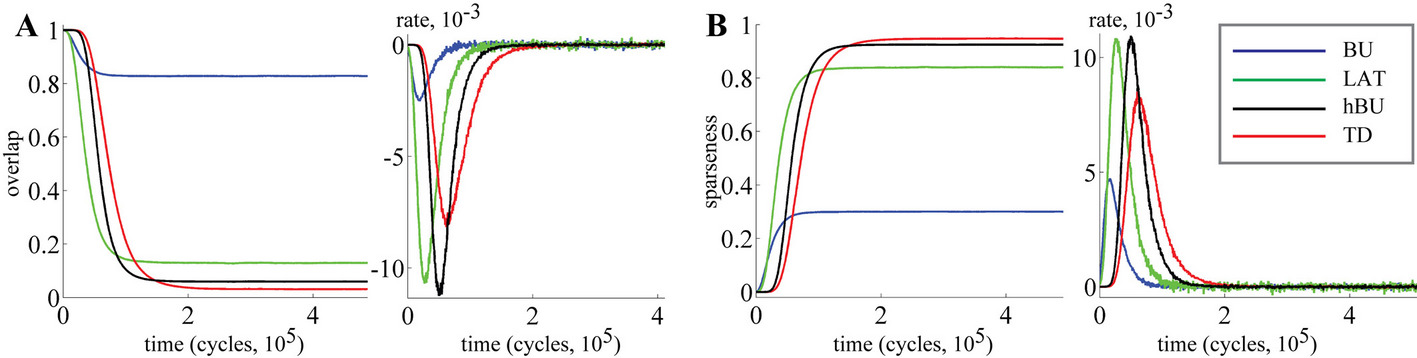}
\caption[Sparseness and overlap]{Overlap {\bf (A)} and sparseness {\bf(B)} time course over $5 \cdot 10^5$ decision cycles for different connectivity types. As the learning progresses, the overlap between the receptive fields is continuously reduced, the connectivity sparseness increases. Again, the temporal sequence of connectivity development is clearly visible (see the text). BU, LAT, hBU, TD denote respectively lower layer bottom-up, lateral, higher layer bottom-up and top-down connectivity types.}
\label{fig:RFSparseOver}
\end{figure}

Each person repeatedly presented to the system during the learning phase leaves a memory trace comprising the parts-based representation of its face on the lower layer and the explicit configurational identity on the higher layer of the memory (Fig. \ref{fig:RFFormation}). The course of gradual differentiation of bottom-up, lateral and top-down connectivity reveals the ongoing process of memory consolidation, where memory traces induced by the face images become more stable and get opportunity to amplify their structure. A common developmental pattern seems to underlie the time courses of structure organization (Sec. \ref{subsec:assess}). There is an initial resting phase, where no structural changes appear, followed by a maturation phase, where massive reorganization occurs and change rate peaks at its maximum value (Fig. \ref{fig:RFDiff}, \ref{fig:RFSparseOver}). Finally a saturation phase is reached, where the structure stabilizes at a certain level of organization and the change rate goes down close to zero. 

Different connectivity types get organized preferentially within a specific time window (Fig. \ref{fig:RFDiff}, \ref{fig:RFSparseOver}). There is a clear temporal sequence of connectivity development, starting with maturation of lower layer bottom up connections, followed by maturation of lateral connections between the bunch columns and by the maturation of bottom-up connectivity of the identity column, ending with the formation of top-down connectivity. Because the development of different connectivity types is highly interdependent, their developmental phases are not disjunct in time, but overlap substantially. In parallel, there is a gradual increase in sparseness within the receptive fields and progressive reduction of the overlap between them. (Fig. \ref{fig:RFSparseOver}) The remaining overlap in associative lateral and configurational bottom-up connectivity reflects the extent to which the parts are shared among different stored face representations.

In the late learning phase, the state of the synaptic structure stabilizes until no substantial changes in the established memory structure can be observed (Fig. \ref{fig:RFDiff}, \ref{fig:RFSparseOver}). Remarkably, the bottom-up connectivity of the bunch columns stays well behind other connectivity types in terms of differentiation grade, sparseness within the receptive fields and their overlap reduction achieved in the final stable state (Fig. \ref{fig:RFDiff}, \ref{fig:RFSparseOver}). While being the latest to initiate its maturation, the top-down connectivity reaches the highest grades of differentiation and sparseness, also being most successful in reducing the overlap. The lateral connectivity between the bunch columns and bottom-up connectivity of the identity column also show comparably high level of organization. These relationships reflect the distinct functional roles the different connectivity types play in their contribution to the memory traces - capturing strongly similar local feature appearance in case of lower layer bottom-up connectivity on the one hand and on the other hand storing weakly overlapping associative and configurational information for different faces in case of lateral and top-down connectivity.

\begin{figure}[!tbhp]
\includegraphics[scale=1.0]{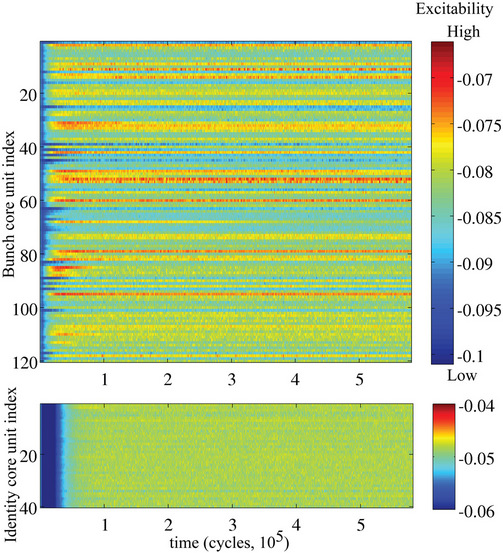}
\caption[Thresholds course]{Time course of excitability regulation. Above the lower, below the higher memory layer. Obvious are the much stronger pronounced differences in excitability between the units on the lower layer.}
\label{fig:Thresholds}
\end{figure}

\begin{figure}[!tbhp]
\includegraphics[scale=1.0]{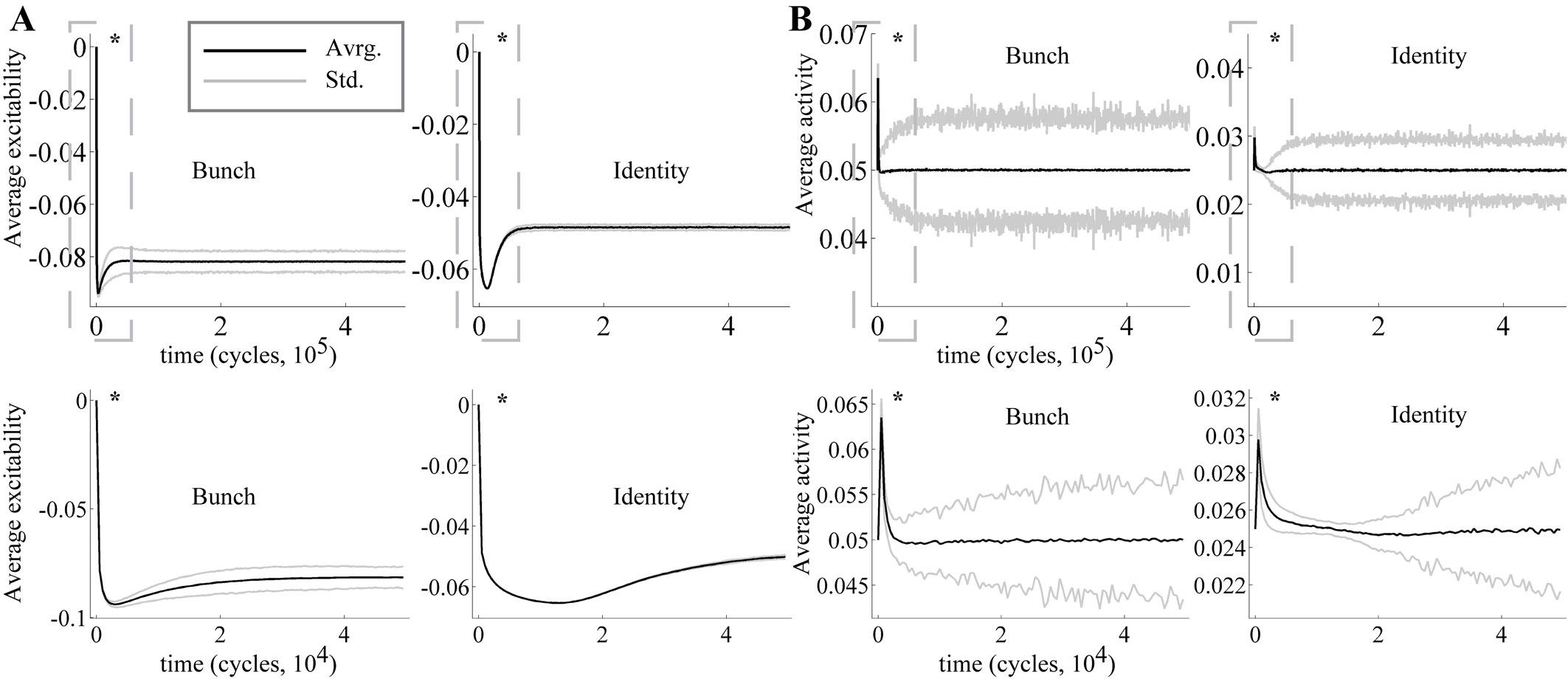}
\caption[Average excitability course]{ {\bf (A)} Time course of average excitability regulation. Above the whole course, below the zoom into down- and upregulation phases. On the left for the bunch units, on the right for the identity units. Black solid curve is the average value, gray curves mark the standard deviation range. The same nomenclature applies for the time course of the average unit activity visualized in {\bf(B)}. The much stronger pronounced differences in excitability between the units on the lower layer are reflected in the greater dispersion of their activities around the average activity level on the lower layer.}
\label{fig:AvgThreshold}
\end{figure}

The changes in the synaptic structure are accompanied by the use-dependent regulation of the excitability thresholds of the core units across the network. Three developmental phases can be distinguished in the time course of excitability modifications (Fig. \ref{fig:Thresholds}). The first phase is characterized by strong and rapid excitability downregulation in the network. This downregulation settles down the core units toward the range of the targeted average activity level $p_{aim}$ ( Eq. \ref{eq:homeostatic}). In this phase, almost no differences between the individual thresholds are present (Fig. \ref{fig:AvgThreshold}). After downregulation crosses its peak, a common upregulation sets in and the differences between the excitability thresholds become much more prominent. The upregulation phase leads to a slight increase of the average excitability and is followed by a saturation phase where the average threshold value stabilizes around certain level. 

Excitability regulation runs differently on different memory layers. On the lower layer the down- and upregulation phases are shorter and occur earlier than the corresponding phases on the higher layer. Moreover, the differences in excitability between the units on the lower layer are much stronger pronounced compared to the rather equalized excitability levels of the higher layer units (Fig. \ref{fig:Thresholds} and \ref{fig:AvgThreshold}).

These differences reflect the distinct functional roles the lower and higher layer play in the memory organization. The lower layer serves as a storehouse for associatively linked distributed facial parts that can be shared by multiple face representations, while the identity units are conjunction-sensitive units representing the configurational identity of the face. Because each memorized person is equally likely to appear on the input, the long-term usage load of the identity units is essentially the same, so no need for a systematic differentiation of excitability thresholds arises there. Part sharing on the other hand imposes different usage frequency on different core units sensitive to different parts, leading to pronounced use-dependent differences in excitability between the bunch column core units.

\subsection{Activity formation and coordination}
\label{subsec:coordination}
The established synaptic structure supports the parts-based representation scheme by encoding the relations between the parts in two alternative ways. First, the relations can be explicitly signaled by the responses of conjunction, or configuration, specific identity core units on the higher layer, each responsible for one of the face identities stored in the memory. Second, the relations can be represented by dynamic assemblies of co-activated part-specific bunch core units, which can be constructed on demand to encode a novel face or to recall an already stored one as a composition of its constituent parts. The selection and binding of the parts-specific and identity-specific units into a coherent assembly coding for an individual face is done in the course of a {\it decision cycle} defined by common unspecific excitatory and inhibitory signals oscillating in the gamma range \citep{Singer1999a, Fries2007}. 

\begin{figure}[!tbhp]
\includegraphics[scale=1.0]{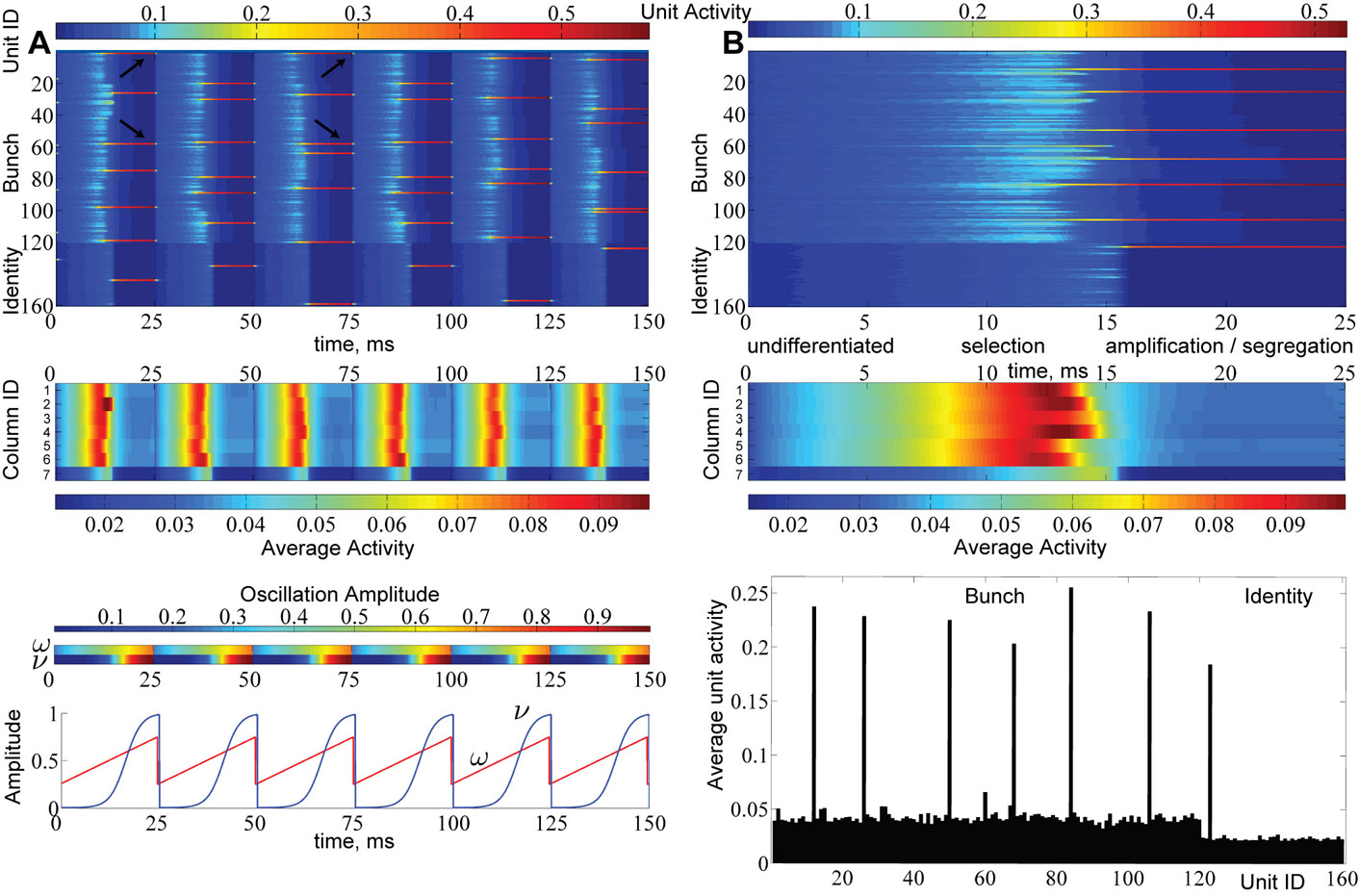}
\caption[Activity course]{Activity formation during the decision cycle. {\bf (A)} A sequence of six successive cycles, each representing a successful recall of a stored individual face. On the top, the activity course is shown, arrows pointing to constituent parts shared by two different face identities. Second and forth cycles show recall of the same face identity. Below is the mean activity course for each column and the oscillation rhythms defining the decision cycle. {\bf(B)} A zoom into a single decision cycle (on the top) to visualize the activity formation phases. Below is the mean activity course for each column and distribution of average unit activities over the decision cycle showing the highly competitive nature of activity formation, where winner units get amplified at the cost of suppressing the others.}
\label{fig:Activity}
\end{figure}

There, the global decision process which may be called binding by competition is responsible for assembly formation, providing clear and unambiguous temporal correlations between the selected units and setting them apart against the rest by amplification of their response strength (Fig. \ref{fig:Activity}). The initial phase of the decision cycle, where the oscillatory inhibition and excitation are low, is characterized by low undifferentiated activities of the network units. With both inhibition and excitation rising, only some of the units are able to resist the inhibition pressure and continue increasing their activity being selected as candidates for assembly formation in the selection phase. Ultimately, the growing competition leads to a series of local winner-take-all decisions across the columns sparsening the activity in the network by strong amplification of a small unit subset at the cost of suppression of the others. In the late phase of a decision cycle, this amplified subset of winner units can be then clearly interpreted as an individual face composed of the local features from respective landmarks and labeled with person's identity, solving the assembly binding problem \citep{CvdM99, Singer1999a}.

A combined view on the mean activity within the columns reveals once more the competitive nature of activity formation in the network (Fig. \ref{fig:Activity}). While the winner unit subset concentrates increasingly high activity, the mean network activation gets progressively reduced at the end of the decision cycle after crossing its peak in the selection phase, indicating that winner subset amplification occurs at the cost of suppressing the rest. Generally, during the whole decision cycle the mean network activity stays at a low level ($p=0.08-0.09$), far below the activity level reached by the winner units subset at the end of the cycle ($p=0.4-0.6$).

One may ask to what extent the competitive activity formation becomes more organized or coherent in terms of representing the memory content as the learning progresses. In other words, we are interested in the level of coherence, or agreement, between the local competitive decisions made in the distributed columns and how it may change with the learning time. One indicator of such coherent behavior is the agreement achieved at the end of the decision cycle between the afferent signals that arrive at network units from different sources such as bottom-up, lateral or top-down. By computing the standard correlation coefficient $\rho$ \citep{DeGroot2001}, we obtain for each afferent signal pair of different sources a course showing the development of the coordination between the signals over the learning time.

\begin{figure}[!bhtp]
\includegraphics[scale=1.0]{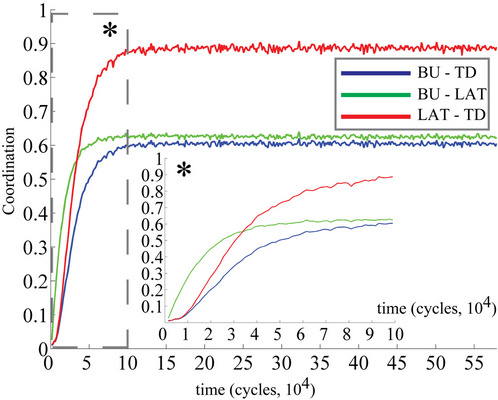}
\caption[Coordination course]{Improvement of signal coordination in the course of learning. Standard correlation coefficients $\rho$ were computed for each signal pair. BU, LAT, TD denote respectively bottom-up, lateral and top-down signals.}
\label{fig:coordination}
\end{figure}

The coordination level between the bottom-up, lateral and top-down signals increases gradually from the initially very low value close to zero toward higher and higher grade (Fig. \ref{fig:coordination}). The low coherence value in the early learning phase reveals the inability of the signals converging on the network units to be in consensus with each other about the local decision outcome, deranging the global decision making. As learning progresses, the signal pathway structure is gradually improved for the storage and representation of the content, leading to stronger and stronger consistency in local signaling. The bottom-up and lateral signals are the first to develop a significant grade of coherence. Slightly later the lateral and top-down signals reach a substantial coherence level and the latest to establish a coordinated cross-talk are the signals from bottom-up and top-down sources. Furthermore, the lateral and top-down signals establish the strongest final grade of coherence that is significantly higher than the coherence between bottom-up and lateral as well as bottom-up and top-down signals. Their coherence still reaches substantial values though, the former being slightly above the latter.

\begin{figure}[!tbhp]
\includegraphics[scale=1.0]{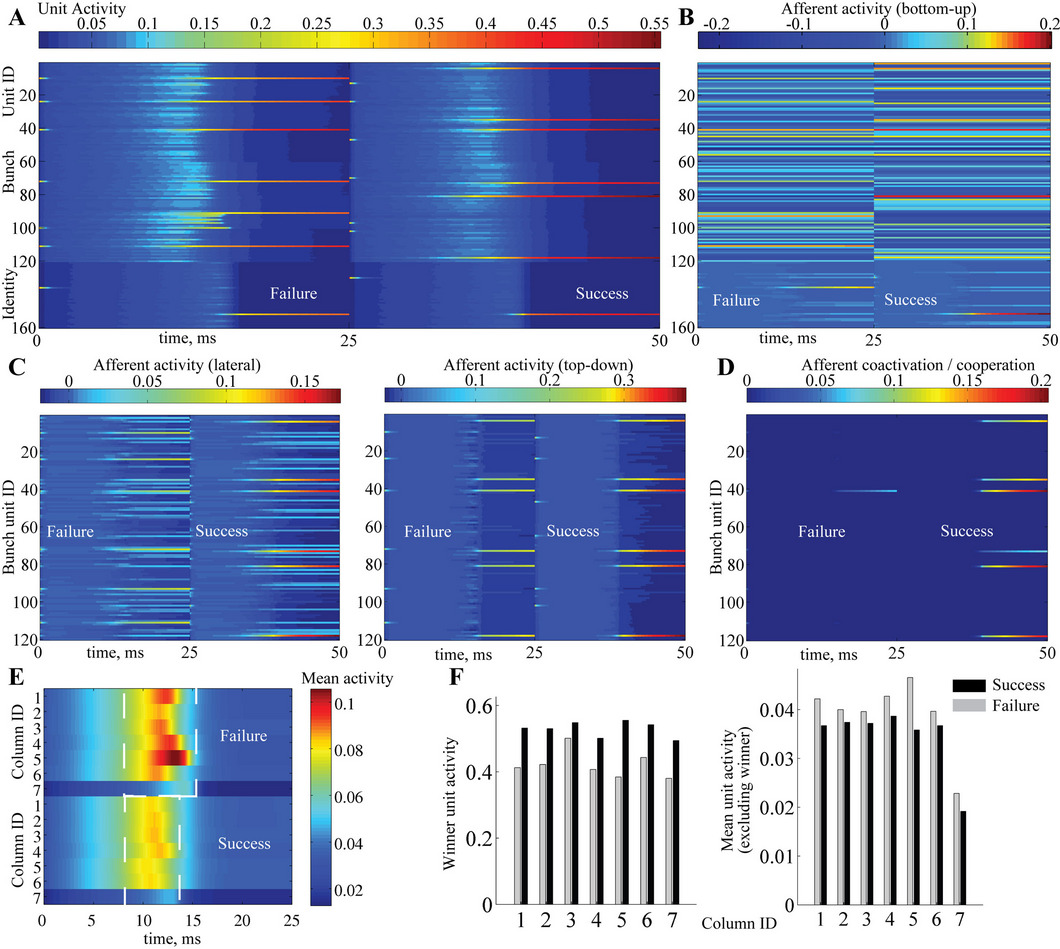}
\caption[Coordination and Recall]{Coordination and activity formation in successful and failed recall. Two decision cycles showing failed and successful recall. {\bf (A)} Network activity course.  {\bf(B)} Bottom-up afferent signals course.  {\bf(C)} Lateral and top-down afferent signals course. {\bf(D)} Signal coordination course assessed by measuring the co-activation of bottom-up, lateral and top-down signals converging on the network units. In the failed recall, there is a clear break-down of signal coordination in afferents converging on the winner units. {\bf(E)} Course of mean activity in the columns. In the failed recall, a substantially increased overall activation is clearly seen as well as the shift of its broader peak to a later time point. {\bf(F)} Winner unit activities at the end of the decision cycle on the left and mean unit activities (excluding the winners) over whole cycle on the right for each column. In the failed recall, winner activities are consistently lower, while the mean rest unit activities are consistently higher than in the successful recall.}
\label{fig:GoodVsBad}
\end{figure}

During the course of a single decision cycle, a co-activation measure can be used to check whether the incoming signals are coordinated properly to make up the decisions. The relationship between the afferent signal coordination and the function of the memory is particularly clear if the coordination level in a successful recall is compared to the coordination shown during a failed recall, where the identity of the person is misclassified (Fig. \ref{fig:GoodVsBad}). In a successful recall, where the facial representation and person's identity are correctly retrieved from the memory, a well-established coordination can be observed between the co-active afferent signals converging on the winner units. In a failed recall, the identity column making a wrong decision sends top-down signals that are not in agreement with the bottom-up and lateral signals conveyed by the bunch columns. As consequence, the signal coordination breaks down, serving as a reliable indicator of a recall failure (Fig. \ref{fig:GoodVsBad} (D)). 

A further indicator that can help in differentiating a successful from a partially or completely failed recall is the activity level of the winner units at the end of the decision cycle. A successful recall is accompanied by a high degree of cooperation between the participating winner units, so that the level of their final activation is high. At the same time, the competitive action of the winner units subset suppresses strongly the rest activity, so that the overall network activity is substantially diminished. Contrarily, a failed recall has something to do with disagreement between some local decisions, resulting in decreased afferent signal coherence, which in turn leads to a much lower level of final activity in the winner units. Their competitive influence is also weakened, leading to a higher overall network activity (Fig. \ref{fig:GoodVsBad} (F)). Thus, a simple comparison of the winner activities to their average level can already provide enough information to conclude about the quality of recall. The recall quality can be assessed on the global level of identity as well as on the component level, where either identity recognition failure or part assignment failure might be stated.

\subsection{Recognition performance}
\label{subsec:recognition}
To assess the recognition capability of the memory, we evaluate the learning and generalization error of two different system configurations. These different configurations, the fully recurrent and purely feed-forward one, are set up to substantiate the hypothesis stating the functional advantage of the recurrent memory structure over the structure with purely feed-forward connectivity. Both configurations were trained under equal conditions and then tested to compare their performance against each other (refer to Sec. \ref{subsec:learning}).

\begin{figure}[!tbhp]
\includegraphics[scale=1.0]{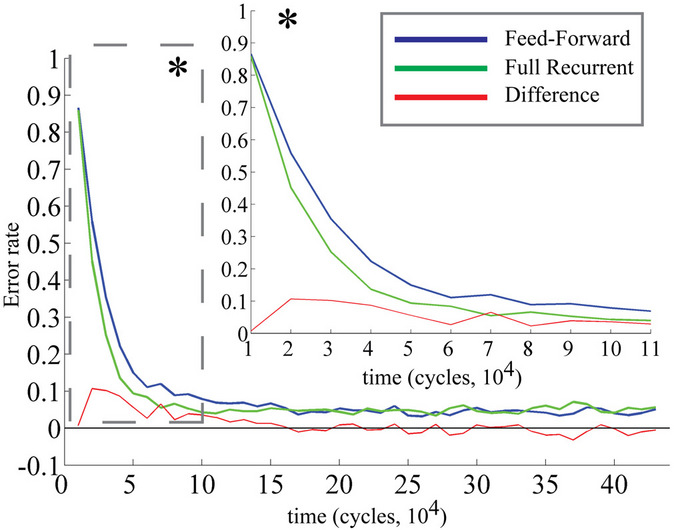}
\caption[Learning error rate]{Learning error rate of feed-forward and fully recurrent memory configuration.}
\label{fig:ErrorRate}
\end{figure}

Both the purely feed-forward and fully recurrent configurations are able to successfully store the facial identities of the persons ($40$ in total) in the memory structure. Strong decay of the learning error over the time is clearly evident for both network configurations. The learning error rate falls rapidly in the early learning phase (first $5 \cdot 10^4$ decision cycles) until it saturates at the values slightly below $5\%$ in the later phase beyond $10^5$ cycles (Fig. \ref{fig:ErrorRate}). Although there is no significant difference in the learning error rate between the both configurations after the saturation level is reached, the time needed to reach the saturation level is substantially shorter for the fully recurrent configuration (saturates around $10^5$ cycles) than for the purely feed-forward one (saturates around $1.5 \cdot 10^5$ cycles). Thus, the learning progresses about $33\%$ faster for the fully recurrent system than for the purely feed-forward one. The fully recurrent configuration seems to speed up the learning progress in the critical early learning phase, probably taking benefit of additional assistance provided by lateral and top-down connectivity for the organization, amplification and stabilization of the memory traces.

At first glance, analysis of the learning error time course suggests that the only functional advantage of the fully recurrent configuration is the learning speed-up observed in the early phase. However, another important functional advantage is revealed if the generalization error rates are compared. The generalization error is measured on the alternative face views not shown during the learning phase (see Tab. \ref{tab:summary}). A striking result is the significant discrepancy in performance between the two configurations manifested on the duplicate views containing emotional expressions (smiling and sad). There, the  error rate difference is about $5\%$ in favor of the fully recurrent memory configuration. The generalization error of purely feed-forward configuration is $38.46\%$ larger on the duplicate smiling view and $62.5\%$ larger on the duplicate sad view than the generalization error of the fully recurrent configuration. On the other views, no significant difference in error rate can be detected between both configurations.

\begin{table}
\subtable{
\begin{tabular}{|l|c c c|}
\hline
\small{Configuration} & \multicolumn{3}{c|}{\small{Views, Error Rate}} \\
\,  & \footnotesize{Original} & \footnotesize{Smiling} & \footnotesize{Sad} \\
\hline
\footnotesize{fully recurrent} & \footnotesize{$0.1\% \pm 0.07\%$} & \footnotesize{$6.06\% \pm 0.58\%$} & \footnotesize{$4.02\% \pm 0.42\%$} \\
\footnotesize{purely feed-forward} & \footnotesize{$0.067\% \pm 0.0528\%$} & \footnotesize{$5.72\% \pm 0.92\%$} & \footnotesize{$3.75\% \pm 0.38\%$} \\
\hline
\end{tabular}
}

\subtable{
\begin{tabular}{|l|c c c|}
\hline
\small{Configuration} & \multicolumn{3}{c|}{\small{Views, Error Rate}} \\
\,  & \footnotesize{Duplicate} & \footnotesize{Duplicate, smiling} &  \footnotesize{Duplicate, sad}  \\
\hline
\footnotesize{fully recurrent} & \footnotesize{$1.64\% \pm 0.16\%$} & \footnotesize{$\mathbf{13.41}\% \pm 0.94\%$} & \footnotesize{$\mathbf{8.74\%} \pm 0.38\%$} \\
\footnotesize{purely feed-forward} & \footnotesize{$1.75\% \pm 0.13\%$} & \footnotesize{$\mathbf{18.42\%} \pm 0.93\%$} & \footnotesize{$\mathbf{13.68\%} \pm 0.64\%$} \\
\hline
\end{tabular}
}
\caption[fully recurrent versus purely feed-forward memory configuration generalization performance]{Comparison of generalization error between the purely feed-forward and fully recurrent memory configuration. The configurations were tested after learning time of $5\cdot10^5$ cycles. fully recurrent configuration shows a significantly better performance on the duplicate views with emotional expressions, while comparable performance is shown on the other views.}
\label{tab:summary}
\end{table}

These results highlight an interesting property of the functional advantage as it has been assessed for the fully recurrent memory configuration. The purely feed-forward configuration falls significantly behind the fully recurrent one only on certain views, performing comparably well on the others. Apparently, the stronger the deviation of the alternative view from the original view showed during the learning, the more evident is the enhancement in generalization capability. Even if given only a short time of a single decision cycle, the recurrent connectivity seems to gain benefit particularly in novel situations, where purely feed-forward processing alone has more difficulties in achieving correct interpretation of the less familiar face view.

%% file: discussion.tex
\section{Discussion}
\label{sec:discussion}
To identify potential neural mechanisms that are responsible for the formation of parts-based representations in visual memory, we examined the process of experience-driven structure self-organization in a model of layered memory. We chose the task of unsupervised open-ended learning and recognition applied to human faces from a database of natural face images. The final goal was to build up a hierarchically organized associative memory structure storing faces of individual persons in a parts-based fashion. Employing slow activity-dependent bidirectional plasticity \citep{Bienenstock1982,Artola1993,Cho2001} together with homeostatic activity regulation \citep{Desai1999b, Zhang2003} and a fast neuronal population dynamics with a strongly competitive nature, the proposed system performed impressively well on the posed task. It demonstrated the ability to simultaneously develop local feature vocabularies and put them in a global context by establishing associative links between the distributed features on the lower memory layer. On the higher layer, the system was able to use the configurational information about relatedness of the sparse distributed features to memorize the face identity explicitly in the bottom-up connectivity of identity units. The captured feature constellations were also projected back to the lower layer via top-down connectivity providing additional contextual support for learning and recognition. The identity recognition performance of the system on the original and alternative face views confirmed the functionality of the established memory structure.

{\bf Generic memory architecture.}\hspace{2mm} When thinking about the processes underlying the memory formation and function, it is remarkable that the structure and activity formation in the model network can be governed by a set of local mechanisms which are the same for all neuronal units and all synapses comprising the network. Saying that they are the same here means that for instance the bidirectional plasticity rule for any synapse has not only the same functional description, but also shares a common set of parameter values such as time constant, etc. This supports the view that the synapses arriving from different origins and contacting their target neuron at different sites of the dendritic tree and soma are a kind of universal learning machines, which may well differ in their impact on the firing behavior of the neuron \citep{Sherman1998,Larkum2004, Friston2005}, while obeying the same generic modification rules. Whether this is indeed the case, is currently a subject of intense debates \citep{Sjoestroem2008, Spruston2008a}. Overall, the organization of the system supports the idea of universal cortical operations involving strong competitive and cooperative effects \citep{Malsburg1988}, which are building up on essentially the same local circuitry and the same plasticity mechanisms utilized in different cortical areas \citep{Mountcastle1997a, Phillips1997, Douglas2004}.

{\bf Competition and cooperation in activity and structure formation.}\hspace{2mm} In our study, it becomes clear that learning itself has to rely on certain important properties of the processing on the fast as well as on the slow time scale. To capture statistical regularities hidden in the local sensory inputs and their global compositions, there have to be mechanisms for selecting and amplifying only a small fraction of available neuronal resources, which then become dedicated to a particular object, specializing more and more for the processing of its local features and their relations. Without proper selection, no learning will succeed. However, without proper learning, no reasonable selection can be expected either. Here, we break this circularity by proposing strong competitive interaction between the units on the fast activity time scale. Given a small amount of neural threshold noise, this interaction is able to break the symmetry of the initial condition due to the bifurcation property of the activity dynamics \citep{Lucke2005}, enforcing the unit selection and amplification in the initial learning phase even in the absence of differentiated structure. The response patterns enforced by competition offer sufficient playground for the learning to ignite and move on to organize and amplify some synaptic structure that is suitable for laying down specific memory content via ongoing slow bidirectional Hebbian plasticity. In combination with competitive activity dynamics, the bidirectional nature of synaptic modification assists further the competition between memory traces as it attempts to reduce the overlap between the patterns the network units preferentially respond to, segregating memory traces in the network structure whenever possible. The state of undifferentiated structure is however the worst-case scenario and not necessarily the initial condition for learning, as there may be basis structures prepared for the representation of many behavioral relevant patterns, like for instance faces \citep{Johnson1991}. Interestingly, the progress from an undifferentiated to a highly organized state via selection and amplification of a small subset of totally available resources is a general feature in evolutionary and ontogenetic development of biological organisms. The notion that the very same principles may guide the activity and structure formation in the brain supports the view of learning as an optimization procedure adapting the nervous structure to the demands put on it by the environment \citep{Malsburg1988, Edelman1993}.

Noteworthy, there is a very important difference in the way how the unit selection, or decision making, is implemented by competition given the early, immature or late, mature state of the connectivity structure. In the immature state where the contextual connectivity is not established yet, the local decisions in the lower layer bunch columns are made completely independent from each other. On the contrary, decision making in the mature state involves interactions between the local decisions via already established lateral and top-down connections. These associative connections enable cooperation within and competition between unit assemblies, promoting a coordinated global decision. The separation of synaptic inputs enables decision making to use information from different origins according to its functional significance - carrying either sensory bottom-up evidence for a local appearance or providing clues for relational binding of distributed parts into a global configuration \citep{Phillips1997}. The agreement between sensory and contextual signaling about the outcome of local decisions improves continuously as learning progresses. The initially independent local decision making becomes thus orchestrated by contextual support formed in the course of previous experience with visual stimuli.

{\bf Signal and plasticity coordination.}\hspace{2mm} The coherency of cooperative and competitive activity formation cannot be guaranteed by the contextual support alone, as the time coordination of decision making across distributed units also matters. The decision cycle, which defines a common reference time window for decision making, orchestrates not only the activities, but also bidirectional synaptic modifications across the units. This reassures that structure modification amplifies the connections within the right subset of simultaneously highly active units encoding a particular face. The cortical processing seems to be reminiscent of oscillatory rhythms in the gamma range used here to model the decision cycle. Particularly, there is evidence that oscillatory activity may serve as reference signal coordinating plasticity mechanisms in cortical neurons \citep{Huerta1995, Wespatat2004}. There is also support for a phase reset mechanism locking the oscillatory activity on the currently presented stimulus \citep{Makeig2002,Axmacher2006}. Taken together, current evidence suggests the possible interpretation of the gamma cycle as a rapidly repeating winner-take-all algorithm as it is modeled in this work \citep{Fries2007}. The winner-take-all competition can be carried out rapidly due to low latencies of fast inhibition and its result can be read out fast (on the scale of few milliseconds) due to the response characteristics of the population rate code \citep{Gerstner2000}.

{\bf Hierarchical parts-based representation.}\hspace{2mm} An essential property of our memory system is parts sharing, as it allows the same basic set of the elementary parts to be used for the combinatorial composition of familiar and novel objects without the need to add new physical units into the system. Endowed with this ability, the memory network can be also interpreted as a layered neuronal bunch graph \citep{CvdM97}, without taking into account the topological information. Here, the graph nodes are columns, each holding a set of features with similar physical (visual appearance) or semantical (category or identity) properties \citep{Tanaka2003}. In such a graph, new object representations can be instantiated in a combinatorial fashion by selecting candidate features from each node. The candidate selection here depends critically on the homeostatic regulation of activity, which reassures that each unit is able to participate in memory formation to an equal extent. By introducing the hierarchy in the graph structure, higher order symbols, like identity of a person, can be explicitly represented by assigning the chosen set of candidate features from the lower memory layer to an identity unit on the higher layer. These higher symbols may be used for a compact representation of exceptionally important persons (VIPs), without discarding the information about their composition which is kept in the top-down connections projecting back to the lower layer. Potentially, it would be also possible to select multiple candidates from a single node, or column, to represent an individual face. Here we use very strong competition leading to a form of activity sparseness termed hard sparseness \citep{Rehn2007}, limiting the number of active units to one per column. While this kind of sparse coding is advantageous for learning individual faces, it may be generally too sparse for representing coarser categories (like male of female). However, the competition strength can in principle be adjusted arbitrarily in a task-dependent manner, either by tuning the core unit gain or by balancing the self-excitation and lateral inhibition. The latter can be easily implemented by altering the amplitude of inhibitory or excitatory oscillations. The alteration could be initiated by some kind of internal cortical signal or state, indicating the task-dependent need for the competition strength. The tuning of the competition strength would allow the formation of less sparse activity distributions, representing the stimulus on a coarser categorical level \citep{Kim2008}.

{\bf Attentional and generative mechanisms in the memory.}\hspace{2mm} Interestingly, contextual lateral and top-down connectivity endows the system with further general capabilities. For instance, selective object-based attention is naturally given in our model, because the priming of the identity units on the higher memory layer by preceding sensory or direct external stimulation would also prime and facilitate the part-specific units on the lower layer via top-down connections, providing them with a clear advantage in the competition against other candidates. This priming can mediate covert attention directed to a specific object, promoting the pop out of its stored parts-based representation while suppressing the rest of the memory content. Generally, the selection and amplification by competition can be interpreted as an attentional mechanism, which focuses the neural resources on processing one object or category at the cost of suppressing the rest \citep{Lee1999a, Reynolds1999a}. Although not exploited in this study, the network model is also able to self-generate activity patterns that correspond to the object representations stored in the memory content in absence of any external input. This ability relies heavily on the lateral and top-down connectivity established by previous experience with visual stimuli, placing the model in remarkable relation to generative approaches explaining construction of data representations in machine learning \citep{Ulusoy2005}. From this perspective, each face identity can be interpreted as a global cause producing the specific activity patterns in the network. The identities are in turn composed of many local causes, i.e.\  their constituent parts. The memory structure captures all the relations between local and global causes, being able to reproduce data explicitly in an autonomous mode. 

{\bf Performance advantage over the purely feed-forward structure.}\hspace{2mm} Finally, we presented sound evidence for the functional advantage of lateral and top-down connectivity over the purely feed-forward structure in the memory formation and recall. First, the recurrent context-based connectivity seems to speed up the learning progress. Second, and at least as essential, recurrent configuration outperforms significantly the purely feed-forward configuration on the test views which deviate strongly from the original views shown during learning. This suggests that contextual processing is able to generalize over new data better than the purely feed-forward solution, which performs comparably on original or only slightly deviating views. This outcome indicates that different processing strategies may prove more useful in different situations. While the recurrent connectivity is mostly beneficial in novel situations, which require additional effort for the interpretation and learning of less familiar stimuli configurations, the feed-forward processing already suffices to do a good and quick job when facing well-known, overlearned situations, where effortful disambiguation is not required due to the strong familiarity of the sensory input. There, the feed-forward processing could benefit from the bottom-up pathway structures formed by previous experience and evoke clear, unambiguous, easily interpretable activity patterns along the processing hierarchy without requiring additional contextual support from lateral and top-down connectivity. There are two predictions arising from this outcome, which can be tested in a behavioral experiment involving subordinate level recognition tasks. First, deactivation of lateral and top-down connectivity in the IT would not change performance for overlearned content, but would impair recognition for less familiar instances of the same stimuli viewed under different conditions, the impairment being the more visible the stronger the viewing condition deviates from the overlearned one. Second, the same deactivation should lead to a measurable decrease in the learning speed, increasing the time needed to reach a certain low level of recognition error.

{\bf Model predictions.}\hspace{2mm} There are some more predictions that can be derived from the system's behavior. One general prediction is that failed memory recall should be accompanied by the higher overall activation along the IT processing hierarchy within the gamma or theta cycle, with the activity of the strongest units at cycle's peak being on contrary diminished. Reversely, a successful recall should be characterized by decreased overall activity in the IT and by increased activity in the winner units cluster. This is also interpretable in terms of signaling the degree of decision certainty, the successful recall being accompanied by greater certainty about the recognition result. Further, a failed recall should involve much more depression (LTD) than potentiation (LTP), a successful recall much more LTP than LTD on the active synapses. In addition, if required to memorize and distinguish very similar stimuli, the recall of such an item should lead to a higher overall activity in the IT network than for items with less similar appearance. The winner units, on contrary, should exhibit a reduced activation due to the inhibition originating from the competing similar content. Again, certainty interpretation of the activity level is possible here: the more similar the stimuli to be discriminated, the lower is the winner activation signaling the decision made, indicating lower certainty about the recognition result. An interesting prediction concerning the bidirectional plasticity mechanism is the erasure of a memory trace after repetitive stimulus-induced recall if LTD/LTP transition threshold is shifted to the higher values, for example due to an artificial manipulation, as performed in experiments of selective memory erasure in mice \citep{Cao2008}.

So far, we provided a demonstration of experience-driven structure formation and its functional benefits in a basic core of what we think can be further developed into a full-featured, hierarchically organized visual memory domain for all kind of natural objects. As usual, several open questions remain, such as invariant or transformation-tolerant processing, development of a full hierarchy from elementary visual features to object categories and identities, establishing the interface for behaviorally relevant context as proposed in the framework of reinforcement learning, incorporating the mechanisms of active vision and so on. Nevertheless, with this work we hope we succeeded not only to highlight the crucial importance of coherent interplay between the bottom-up and top-down influences in the process of memory formation and recognition, but also to gain more insight into the basic principles behind the self-organization \citep{Malsburg2003} of a successful subsystem coordination across different time scales. Aiming for real world applications, we believe that the incremental, unsupervised open-ended learning design instantiated in this work provides an inspiring and guiding paradigm for developing systems capable of discovering and storing complex structural regularities from natural sensory streams over multiple descriptional levels.

%% file: acknowledge.tex
\section*{Disclosure/Conflict-of-Interest Statement}
\label{sec:disclosure}
The authors declare that the research was conducted in the absence of any commercial or financial relationships that could be construed as a potential conflict of interest.

\section*{Acknowledgments}
\label{sec:acknowledge}
We would like to thank Cristina Savin, Cornelius Weber and Urs Bergmann for the helpful corrections on this manuscript. This work was supported by the EU project DAISY, FP6-2005-015803.